\documentclass{article}
\usepackage{amsmath}
\usepackage{amsthm}
\usepackage{graphicx}
\usepackage[tight,footnotesize]{subfigure}
\usepackage{epstopdf}
\usepackage{booktabs}
\usepackage{threeparttable}
\usepackage{float}
\graphicspath{{./figures/}}
\begin{document}

\title{New Variational Method of Grid Generation with Prescribed Jacobian Determinant and Prescribed Curl}
\author{Xi Chen, Guojun Liao}
\date{}
\maketitle

\begin{abstract}
Adaptive grid generation is an active research topic for numerical solution of differential equations. In this paper, we propose a variational method which generates transformations with prescribed Jacobian determinant and curl. Then we use this transformation to achieve adaptive grid generation task.
\end{abstract}
{\bf Keywords:} Grid Generation, Variational Method, Jacobian determinant, Curl, Reconstruction
\section{Introduction}
The deformation method \cite{Feng98},\cite{Liao00},\cite{Liao09} has its origin in differential geometry \cite{Moser65}. Its main theoretical advantages is that the grid transformation generated has prescribed Jacobian determinant. Its implementation is based on calculating a velocity field from a differential system consisting of the divergence operator and the curl operator. In the special case when the curl vector is set to zero, the differential system reduces to three (decoupled) scalar Poisson equations in 3D. Among existing methods are several interesting variational methods \cite{Brack82},\cite{Castillo88},\cite{Huang03}, which enjoy clear physical interpretations and are straight forward to implement. The problem we study in this paper is to construct numerically a transformation $\textbf{T}$ on a two dimensional domain $\Omega$, such that, the Jacobian determinant of $\textbf{T}$, namely $J(\textbf{T})$, equals a prescribed monitor function $f_0$, and the curl of $\textbf{T}$, $curl(\textbf{T})$, equals a prescribed monitor function $g_0$:
\begin{equation}
\begin{aligned}
J(\textbf{T}(\mathbf{x}))&=f_0(\mathbf{x})\\
curl(\textbf{T}(\mathbf{x}))&=g_0(\mathbf{x})
\end{aligned}
\end{equation}
We define the $L^2-norm$ of the difference between $J(\textbf{T})$ and $f_0$, $curl(\textbf{T})$ and $g_0$ as
\begin{equation}
ssd=\frac 1 2 \iint_{\Omega} [(J(\textbf{T})-f_0)^2+\alpha (curl(\textbf{T})-g_0)^2] dA
\end{equation}
where $\alpha > 0$ is a weight parameter.\\
Our goal is to construct the desired transformation $\textbf{T}$ by minimizing $ssd$, with respect to some control function $\textbf{f}$ under the following constraints:
\begin{equation}
\textbf{T}(\mathbf{x}) =\mathbf{x} + \textbf{u}(\mathbf{x})
\end{equation}
Where the displacement $\mathbf{u}$ satisfies the Poisson's equation
\begin{equation}
\Delta \textbf{u} = f \\
\end{equation}
with fixed boundary conditions. Given arbitrary $\delta \textbf{f}$, we have $\delta \textbf{f} = \delta \Delta \textbf{u} = \Delta (\delta \textbf{u})$. So $\delta \textbf{u}$ is governed by $\delta \textbf{f}$ through a Poisson's equation.\\
In order to use the gradient descent method to minimize $ssd$ with respect to control function $\textbf{f}$, we shall derive the gradient $\frac{\partial ssd}{\partial \textbf{f}}$ based on the fixed Dirichlet boundary condition. Then we will demonstrate the method by examples of recovering some known transformations.

\section{Derivation of the Gradient}
We focus on the case where the grid nodes on the boundary are fixed. Since the boundary nodes are fixed, the method enjoys convenience in computation and elegant derivation in mathematics. It turns out the general cases where boundary nodes are moved, can be reduced to this fixed boundary case, see Remark 1 at the end of this section.\\
Now we let $\textbf{T}=(T_1,T_2), \textbf{u}=(u_1,u_2), \textbf{f}=(f_1,f_2)$, and take $\alpha =1$ for the simplicity of presentation. For the general cases, see Remark 2 at the end of this section. So,
\[
ssd = \frac 1 2 \iint_{\Omega}[(J(\textbf{T})-f_0)^2 + (curl(\textbf{T})-g_0)^2] dA
\]
and the displacement $\textbf{u}$ satisfies now,
\begin{equation}
\left\{
\begin{array}{lll}
\Delta u_1 &=f_1\\
\Delta u_2 &=f_2\\
u_1,u_2&=0 \qquad on \ \partial \Omega
\end{array}
\right.
\end{equation}
By variational calculus, we have
\[
\begin{array}{llll}
\delta ssd &=\iint_{\Omega}(J(\textbf{T})-f_0)\delta J(\textbf{T}) + (curl(\textbf{T})-g_0)\delta curl(\textbf{T}) dA\\
&=\iint_{\Omega}[(J(\textbf{T})-f_0)\delta (T_{1x}T_{2y}-T_{1y}T_{2x}) + (curl(\textbf{T})-g_0)\delta (T_{2x}-T_{1y})]dA \\
&=\iint_{\Omega}[(J(\textbf{T})-f_0)(\delta T_{1x}T_{2y}+T_{1x}\delta T_{2y}-\delta T_{1y}T_{2x}-T_{1y}\delta T_{2x})\\ &+ (curl(\textbf{T})-g_0)(\delta T_{2x}-\delta T_{1y})]dA
\end{array}
\]
From (3), we know $\delta \textbf{T} = \delta \textbf{u}$, or $\delta T_1 =\delta u_1, \delta T_2 =\delta u_2$, so $\delta T_{ix} =\delta u_{ix}, \delta T_{iy} =\delta u_{iy}, i=1,2$, plug back in the above equation, we can continue as:
\[
\begin{aligned}
\delta ssd &= \iint_{\Omega}[(J(\textbf{T})-f_0)(\delta u_{1x}T_{2y}+T_{1x}\delta u_{2y}-\delta u_{1y}T_{2x}-T_{1y}\delta u_{2x}) \\ &+ (curl(\textbf{T})-g_0)(\delta u_{2x}-\delta u_{1y})]dA
\end{aligned}
\]
Let $P=(J(\textbf{T})-f_0), Q=(curl(\textbf{T})-g_0)$, we have
\[
\delta ssd = \iint_{\Omega}[P(\delta u_{1x}T_{2y}+T_{1x}\delta u_{2y}-\delta u_{1y}T_{2x}-T_{1y}\delta u_{2x})+Q(\delta u_{2x}-\delta u_{1y})] dA
\]
Now, notice that $\nabla \delta u_i =(\delta u_{ix}, \delta u_{iy}), i=1,2$
\[
\begin{aligned}
\delta ssd = \iint_{\Omega}[P((T_{2y},-T_{2x})\cdot \nabla \delta u_1 +(-T_{1y},T_{1x})\cdot \nabla \delta u_2) \\
+ Q((0,-1)\cdot \nabla \delta u_1 + (1,0)\cdot \nabla \delta u_2)]dA
\end{aligned}
\]
Let
\begin{equation}
\left\{
\begin{array}{ll}
-\textbf{a}_1 &= P(T_{2y},-T_{2x})+Q(0,-1)\\
-\textbf{a}_2 &= P(-T_{1y},T_{1x})+Q(1,0)
\end{array}
\right.
\end{equation}
we get
\[
\delta ssd = \iint_{\Omega}(-\textbf{a}_1\cdot \nabla \delta u_1 - \textbf{a}_2\cdot \nabla \delta u_2)dA
\]
Before further proof, we need to list 2 corollaries of divergence theorem which will be used later.
\newtheorem{corollary}{Corollary}
\begin{corollary}
By Applying the divergence theorem to the product of a scalar function $v$ and a vector field \textbf{u} on a 2D domain $\Omega$, we have
\[
\iint_{\Omega}(\textbf{u}\cdot \nabla v + v(\nabla \cdot \textbf{u})) dA = \int_{\partial \Omega}(v\textbf{u}\cdot \textbf{n})dS
\]
where \textbf{n} is the normal vector.
\end{corollary}
\begin{corollary}
By Applying the divergence theorem to the product of a scalar function $v$ and a vector field $\nabla u$ on a 2D domain $\Omega$, we have
\[
\iint_{\Omega}(\nabla u\cdot \nabla v + v(\Delta u)) dA = \int_{\partial \Omega}(v\nabla u\cdot \textbf{n})dS
\]
where \textbf{n} is the normal vector.
\end{corollary}
Notice in Poisson equation (5), we have $u_i=0$ on $\partial \Omega$, so $\delta u_i=0$ on $\partial \Omega$. By Corollary 1 we can continue as
\[
\begin{array}{lll}
\delta ssd &= \iint_{\Omega}(-\textbf{a}_1\cdot \nabla \delta u_1 - \textbf{a}_2\cdot \nabla \delta u_2)dA\\
&= \iint_{\Omega}((\nabla \cdot \textbf{a}_1)\delta u_1 + (\nabla \cdot \textbf{a}_2)\delta u_2)dA - \int_{\partial \Omega}(\delta u_1(\textbf{a}_1 \cdot \textbf{n}) + \delta u_2(\textbf{a}_2 \cdot \textbf{n}))dS\\
&= \iint_{\Omega}((\nabla \cdot \textbf{a}_1)\delta u_1 + (\nabla \cdot \textbf{a}_2)\delta u_2)dA
\end{array}
\]
Now Let $\textbf{g}=(g_1,g_2)$ satisfies
\begin{equation}
\left \{
\begin{array}{lll}
\Delta g_1 &= \nabla \cdot \textbf{a}_1\\
\Delta g_2 &= \nabla \cdot \textbf{a}_2\\
g_1,g_2&=0 \qquad on \ \partial \Omega
\end{array}
\right.\\
\end{equation}
then we have now,
\[
\delta ssd = \iint_{\Omega}(\Delta g_1\delta u_1 + \Delta g_2\delta u_2)dA
\]
By the Corollary 2, for $i=1,2$
\[
\left\{
\begin{array}{lll}
\iint_{\Omega}(\nabla g_i \cdot \nabla \delta u_i + \delta u_i\Delta g_i)dA &=\int_{\partial \Omega}(\delta u_i\nabla g_i\cdot \textbf{n})dS &=0 \\
\iint_{\Omega}(\nabla g_i \cdot \nabla \delta u_i + g_i \Delta \delta u_i)dA &=\int_{\partial \Omega}(g_i\nabla \delta u_i \cdot \textbf{n})dS &=0
\end{array}
\right.
\]
along with (5), finally we will have
\[
\begin{array}{lll}
\delta ssd &= \iint_{\Omega}(\Delta g_1\delta u_1 + \Delta g_2\delta u_2)dA\\
&= \iint_{\Omega}(g_1\delta (\Delta u_1) + g_2\delta(\Delta u_2))dA\\
&= \iint_{\Omega}(g_1\delta f_1 + g_2\delta f_2)dA
\end{array}
\]
which is
\begin{equation}
\frac{\partial ssd}{\partial f_i} =g_i, i=1,2
\end{equation}
\newtheorem{Rmk}{Remark}
\begin{Rmk}
At the end of proof, we want to point out that if we change the identity map $\mathbf{x}$ to any given transformation $\mathbf{T^{*}}(\mathbf{x})$ in (3), the derivation above is still right. Generally speaking, (3) can be replaced by
\[
\textbf{T}(\mathbf{x}) =\mathbf{T^{*}}(\mathbf{x}) + \textbf{u}(\mathbf{x}) \tag{3$^*$}
\]
where $\mathbf{T^{*}}(\mathbf{x})$ is any given transformation.
\end{Rmk}
\begin{Rmk}
For arbitrary given weight parameter $\alpha$ in (2), the derivation above is still right by letting $Q=\alpha (curl(\textbf{T})-g_0)$
\end{Rmk}

\section{Implementation and Experiments}
\subsection{Implementation by the Gradient Descent method}
Now, we can implement our method by gradient descent method as follows:
\begin{enumerate}
  \item \textbf{Initialize} $\mathbf{T}=\mathbf{id}, \mathbf{u}=\mathbf{0},f=\mathbf{0}$
  \item \textbf{Compute} $\mathbf{a}_1,\mathbf{a}_2$ by (6), then \textbf{solve} Poisson's equation (7) to get $g_1,g_2$
  \item \textbf{Update} $f$ by $f_{i,new}=f_{i,old}-g_i\times tstep$, where $tstep$ is an optimization parameter
  \item \textbf{Solve} Poisson's equation (5) to get $u_1,u_2$
  \item \textbf{Update} $\mathbf{T}$ by (3)
  \item \textbf{Back} to 2, iterate until a preset tolerance or a preset number of iteration steps is reached
\end{enumerate}

\subsection{Experiments of Recovering Transformations}
For the purpose of testing the accuracy of our method, and discovering more details inside, we design a recovering experiment. By the derivation, if given a known transformation $\mathbf{T}_0$, we can compute its Jacobian determinant and curl as our $f_0,g_0$, namely $f_0(x)=J(\textbf{T}_0(x)), g_0(x)=curl(\textbf{T}_0(x))$, then the desired transformation $\mathbf{T}$ constructed here should equal $\mathbf{T}_0$. The following figures and tables illustrate corresponding result. In order to compare, we also give the cases when only consider the Jacobian term in our method. We can see clearly the curl plays an important role in grid transformation, and with both the Jacobian and curl term here in our algorithm, we can recover a transformation, which proves the algorithm is reliable and accurate in grid generation.
\subsubsection{Recover a transformation $\textbf{T}_0$ with fixed boundary nodes}
Given a nonlinear transformation $\textbf{T}_0$ from the square $[1,65] \times [1,65]$ to itself, we want to reconstruct $\textbf{T}_0$ from its jacobian determinant and curl. We assume that the boundary nodes of $\textbf{T}_0$ coincide with the nodes of the background grid. We have the following results for $\alpha=1,0.1,10$ respectively. And for different values of $\alpha$, there is no significant difference.
\begin{figure}[H]
\begin{center}
\subfigure[][Only Jacobian]{\includegraphics[width=0.32\textwidth]{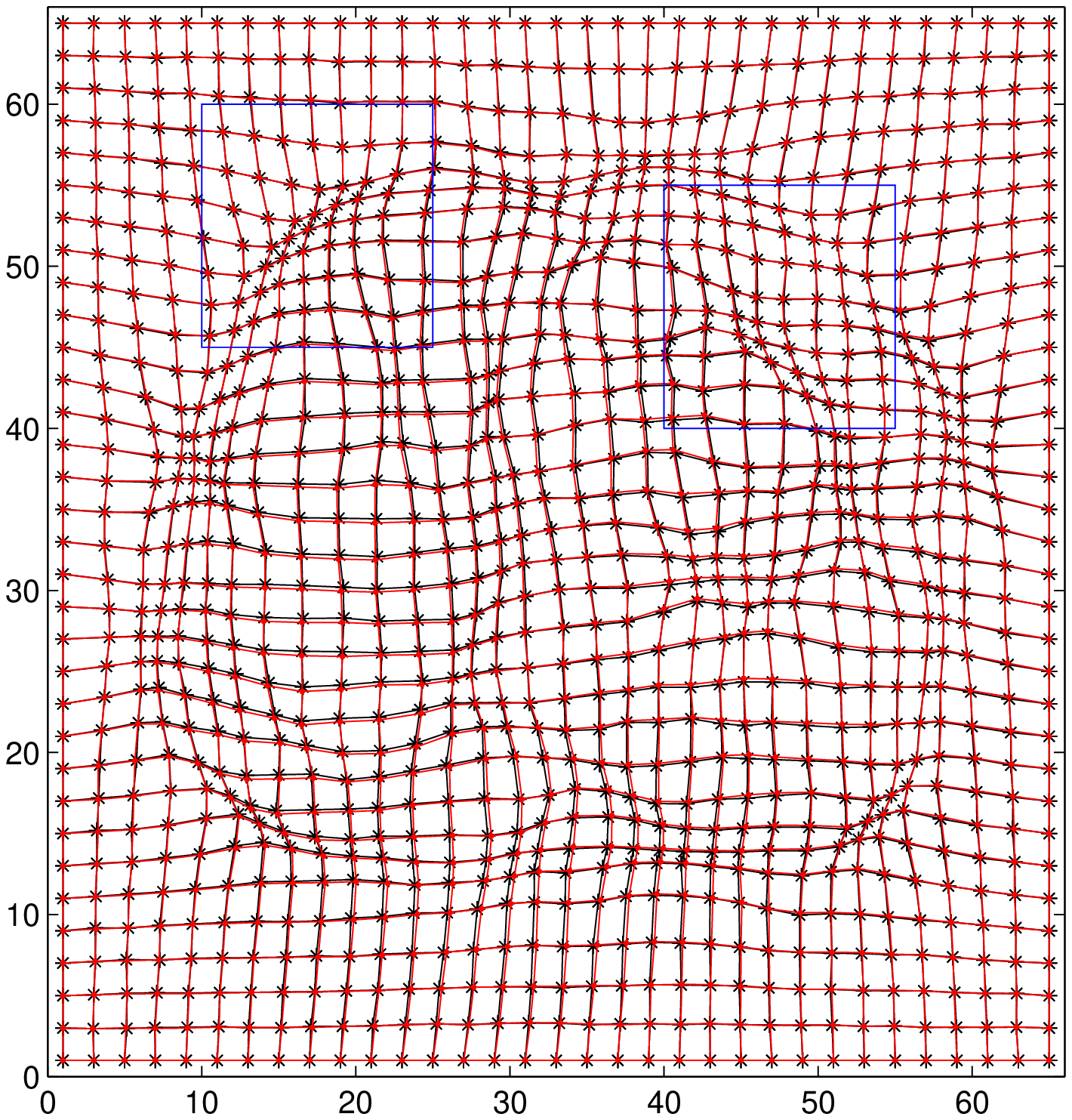}}
\subfigure[][Only Jacobian, enlarged view of left rectangle]{\includegraphics[width=0.32\textwidth]{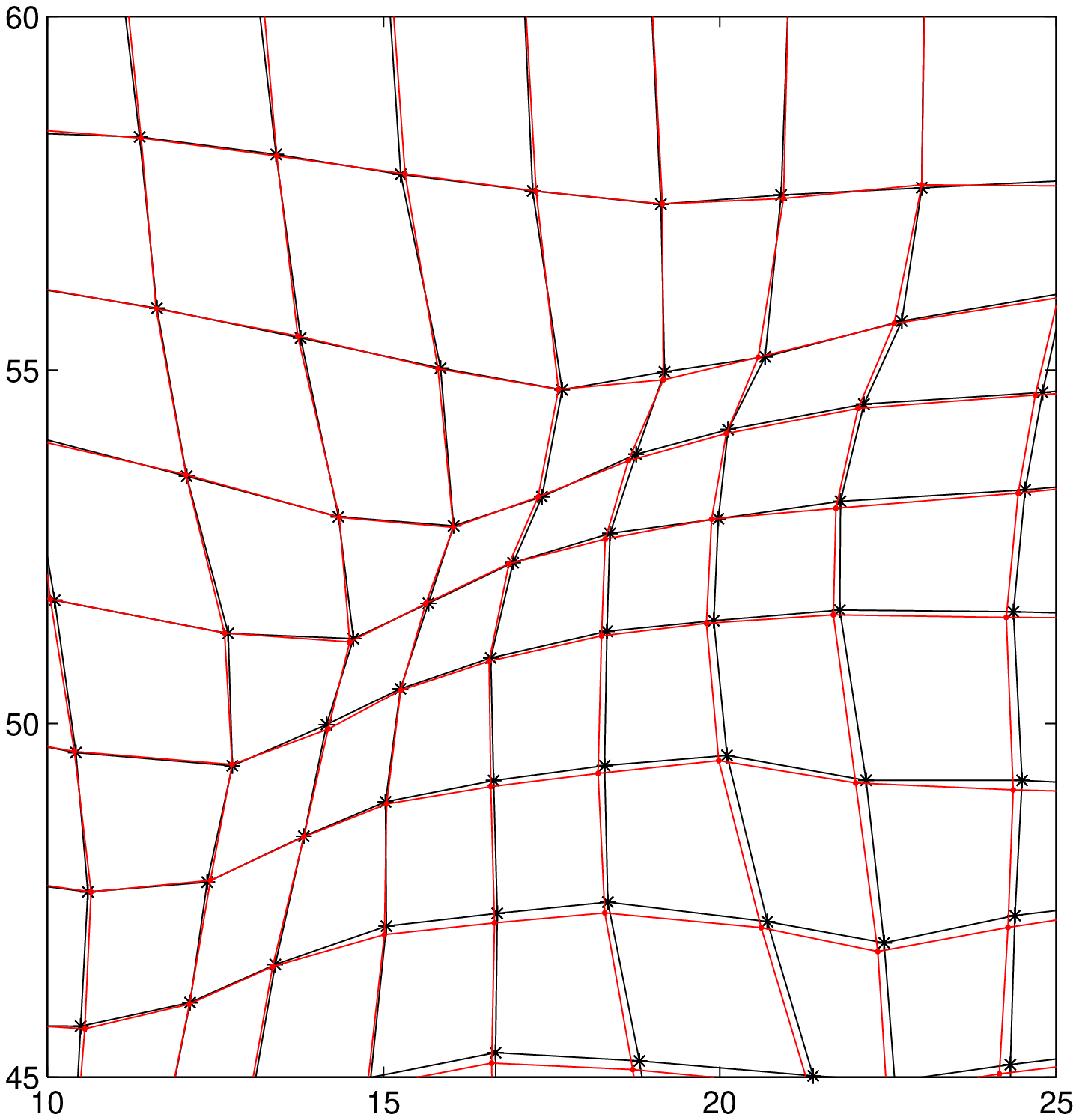}}
\subfigure[][Only Jacobian, enlarged view of right rectangle]{\includegraphics[width=0.32\textwidth]{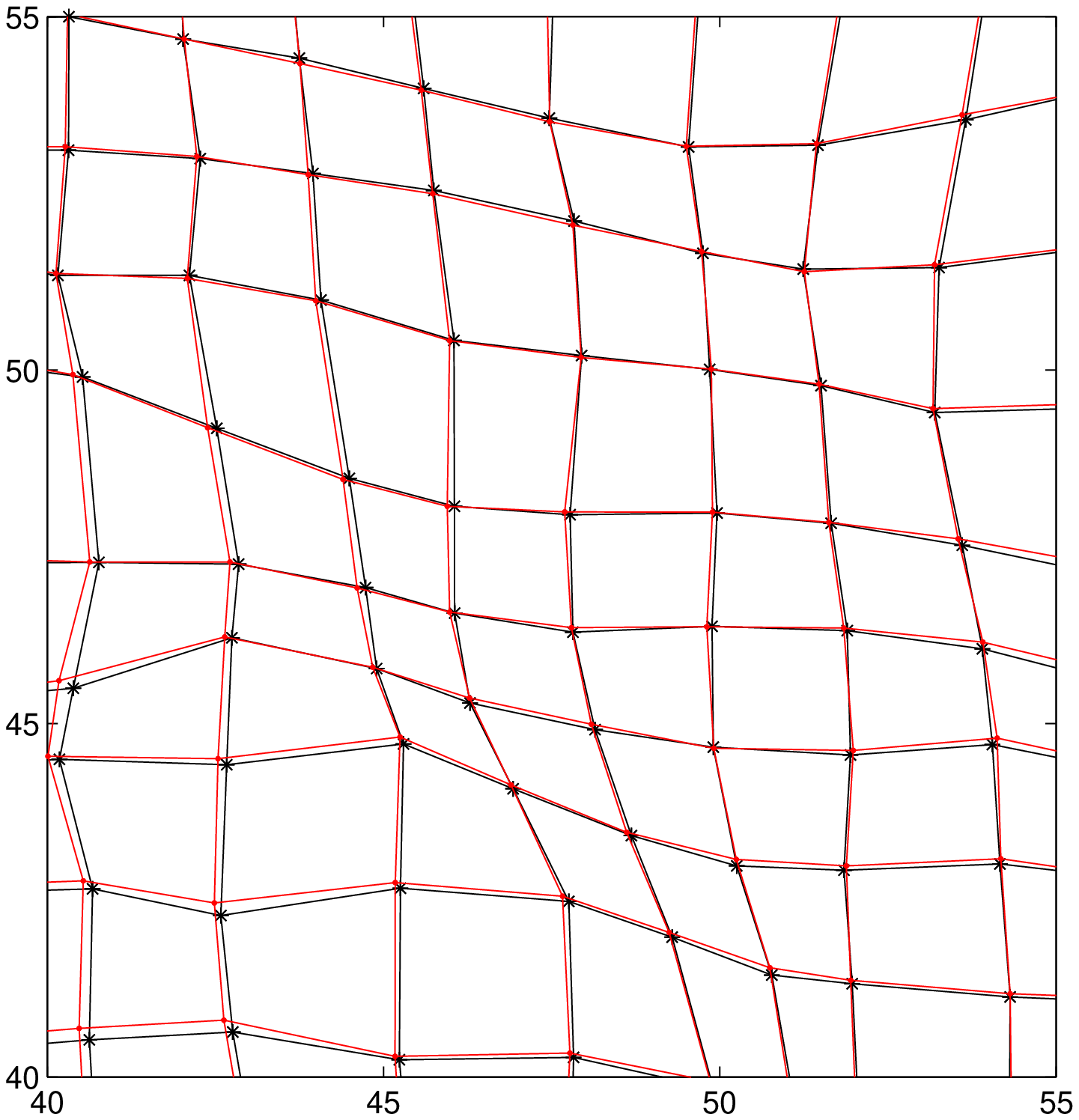}}
\subfigure[][Jacobian and Curl]{\includegraphics[width=0.32\textwidth]{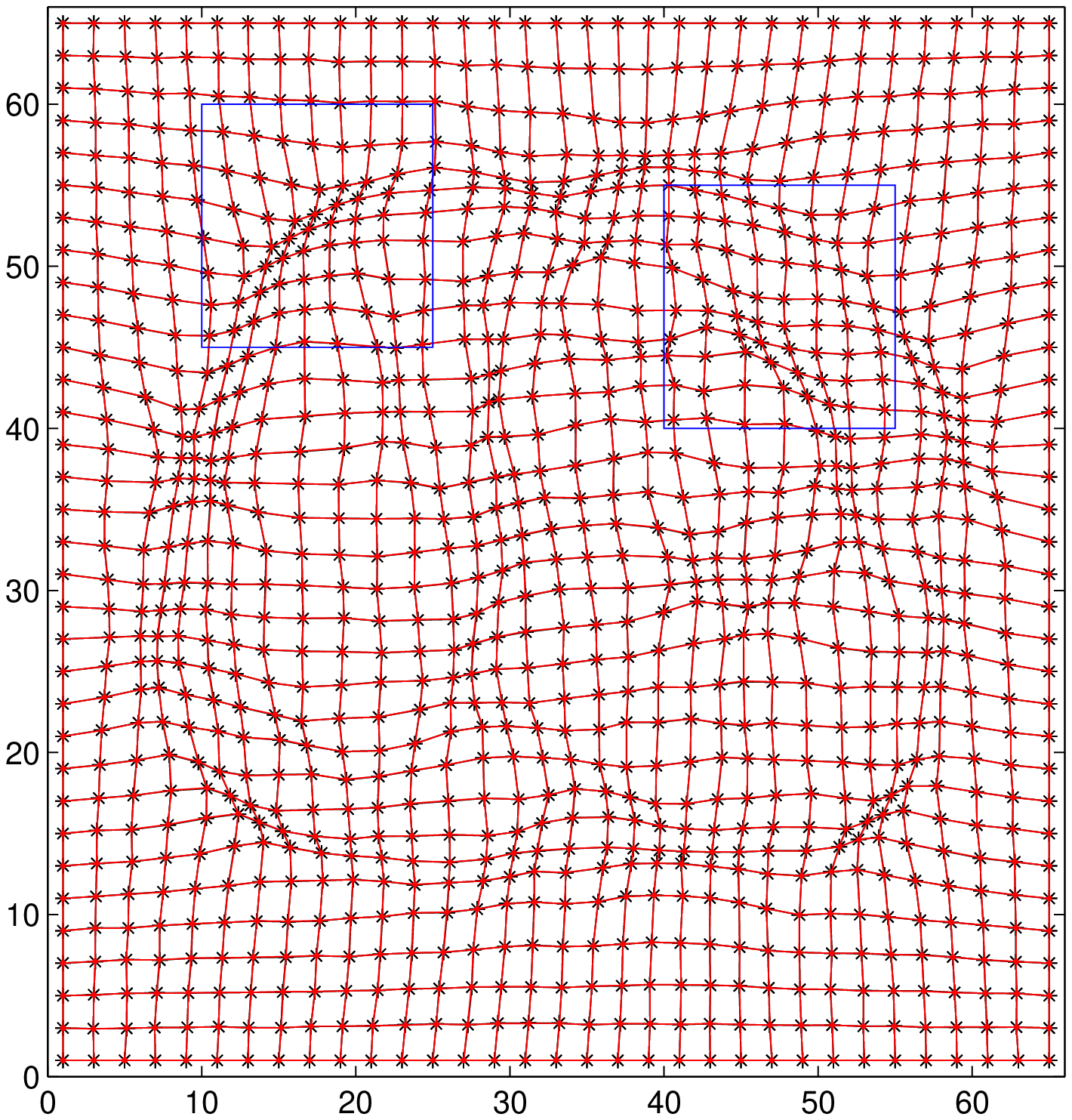}}
\subfigure[][Jacobian and Curl, enlarged view of left rectangle]{\includegraphics[width=0.32\textwidth]{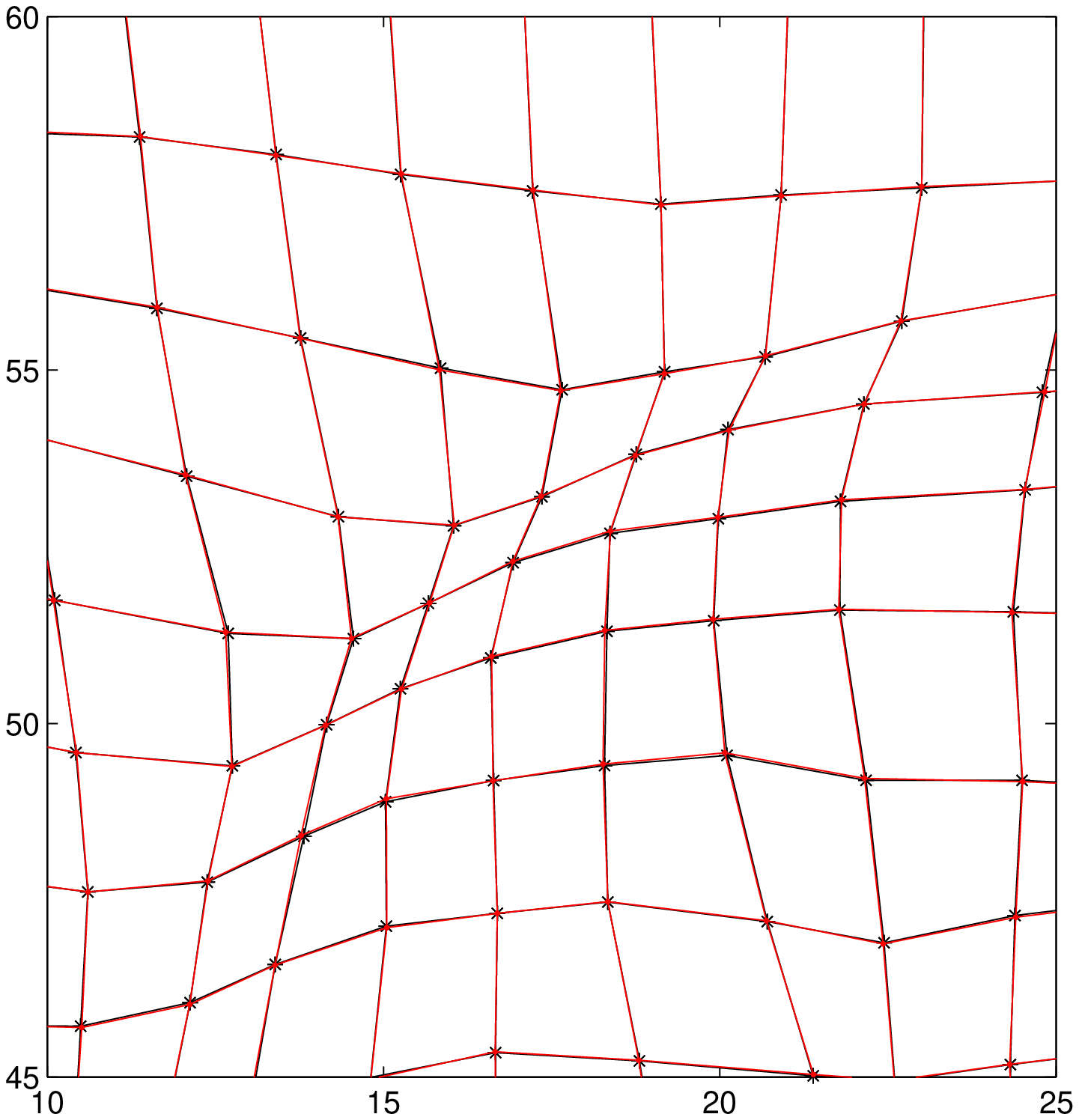}}
\subfigure[][Jacobian and Curl, enlarged view of right rectangle]{\includegraphics[width=0.32\textwidth]{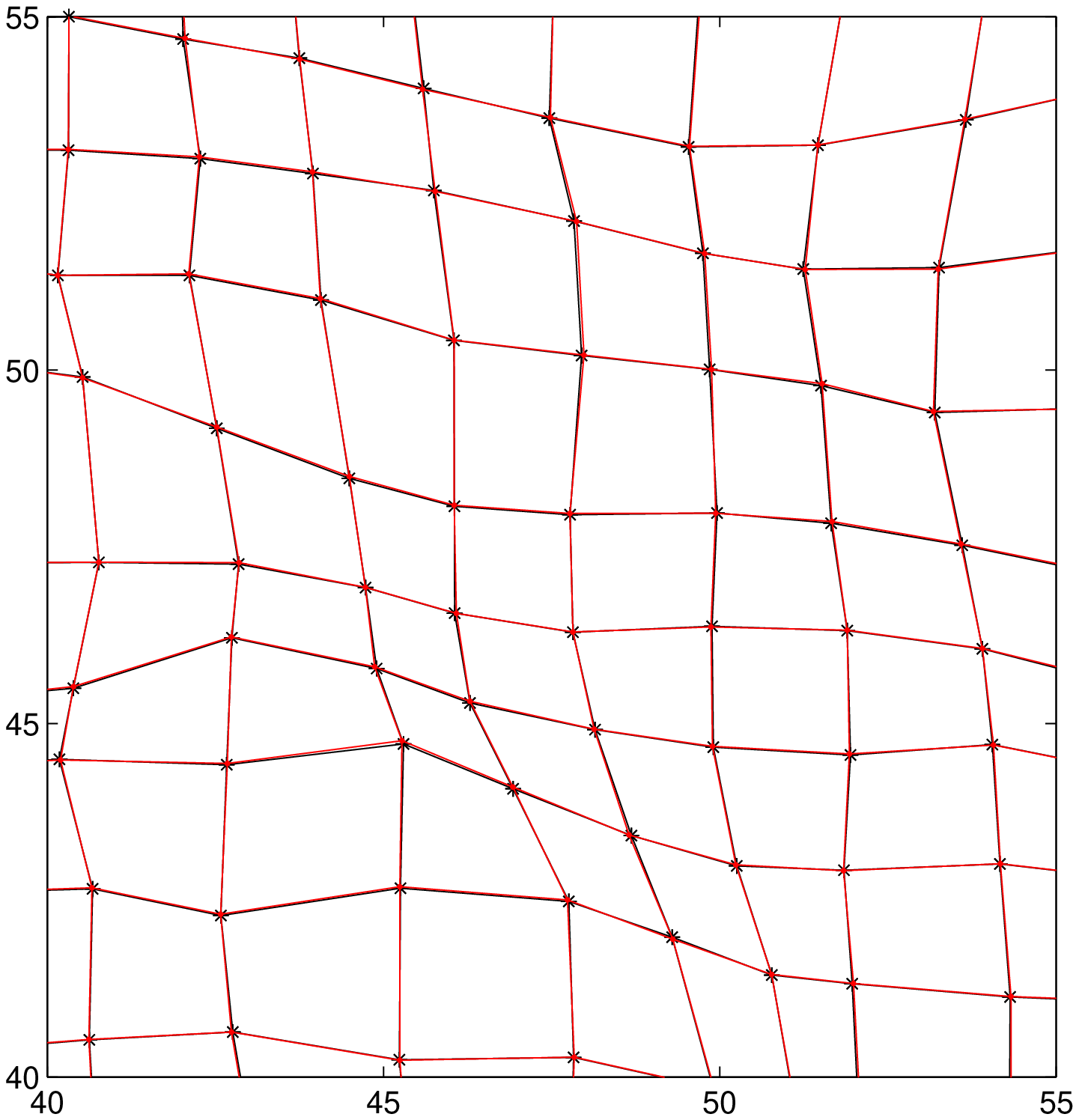}}
\caption{Experiment 1.1, $\alpha=1$, $65 \times 65$ grid nodes, 2000 iteration steps, 9.12sec by Matlab on a Dell 780 PC.  \footnotesize{The black star dots $*$ represent $\mathbf{T}_0$, and red dots $\cdot$ represent constructed $\mathbf{T}$}}
\end{center}
\end{figure}

\begin{table}[H]
\begin{center}
\begin{threeparttable}
\begin{tabular}{|l|c|c|}\hline
    & \emph{Only Jacobian} & \emph{Jacobian and Curl} \\ \hline
    \emph{$ssd_J$$^{1}$} &0.2640 &0.3301 \\ \hline
    \emph{ssd} & 4.8134 & 0.6377 \\ \hline
    \emph{maximal distance$^{2}$} & 0.3444 & 0.0706 \\ \hline
    \emph{average distance} & 0.1109 & 0.0177 \\ \hline
    \emph{maximal angle difference$^{3}$} & 25.8343 & 15.7230 \\ \hline
    \emph{average angle difference} & 2.6132 & 1.7481 \\ \hline
\end{tabular}
\small
\begin{tablenotes}
    \item [1] $ssd_J$ means only consider Jacobian part in $ssd$, $^2$distance means $\|\textbf{T}(\mathbf{x})-\textbf{T}_0(\mathbf{x})\|_{2}, \mathbf{x} \in \Omega$, $^3$angle difference means the corresponding angle differences in every corresponding quadrilateral
\end{tablenotes}
\end{threeparttable}
\caption{Comparison of Experiment 1.1}
\end{center}
\end{table}

\begin{figure}[H]
\begin{center}
\subfigure[][Only Jacobian]{\includegraphics[width=0.32\textwidth]{recoverJplusnoisesPhi6Fbrain_1_1.eps}}
\subfigure[][Only Jacobian, enlarged view of left rectangle]{\includegraphics[width=0.32\textwidth]{recoverJplusnoisesPhi6Fbrain_1_2.eps}}
\subfigure[][Only Jacobian, enlarged view of right rectangle]{\includegraphics[width=0.32\textwidth]{recoverJplusnoisesPhi6Fbrain_1_3.eps}}
\subfigure[][Jacobian and Curl]{\includegraphics[width=0.32\textwidth]{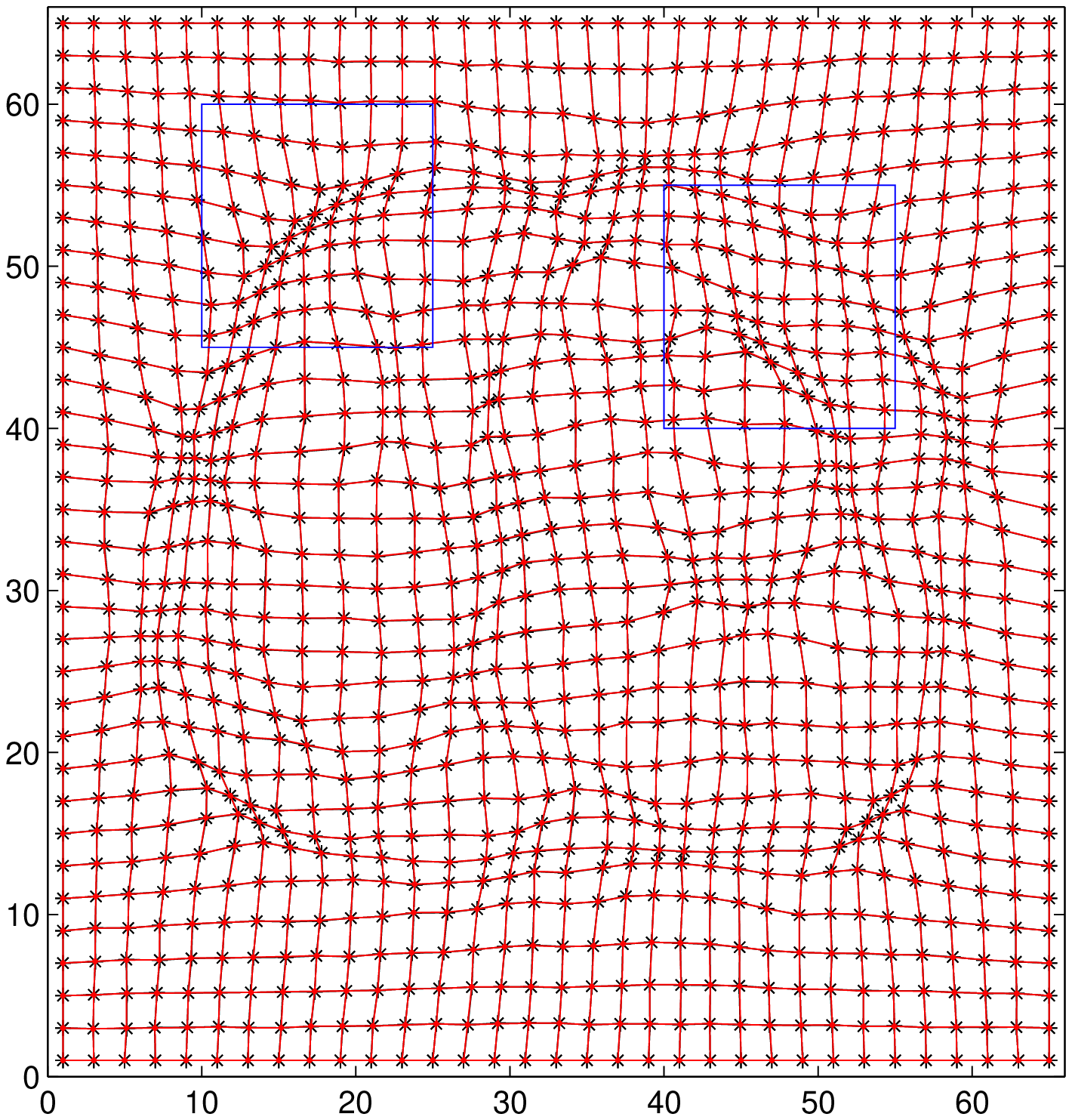}}
\subfigure[][Jacobian and Curl, enlarged view of left rectangle]{\includegraphics[width=0.32\textwidth]{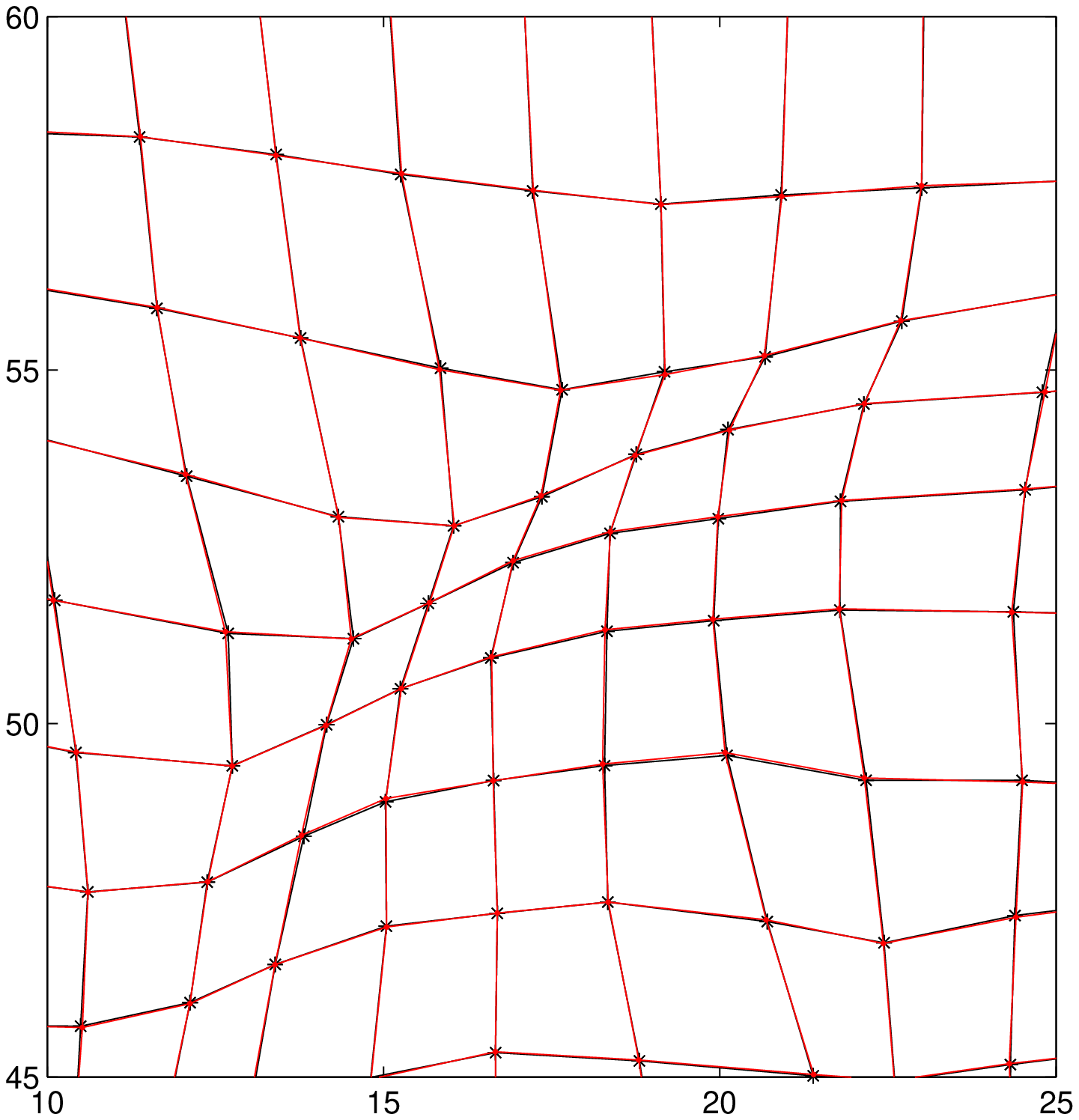}}
\subfigure[][Jacobian and Curl, enlarged view of right rectangle]{\includegraphics[width=0.32\textwidth]{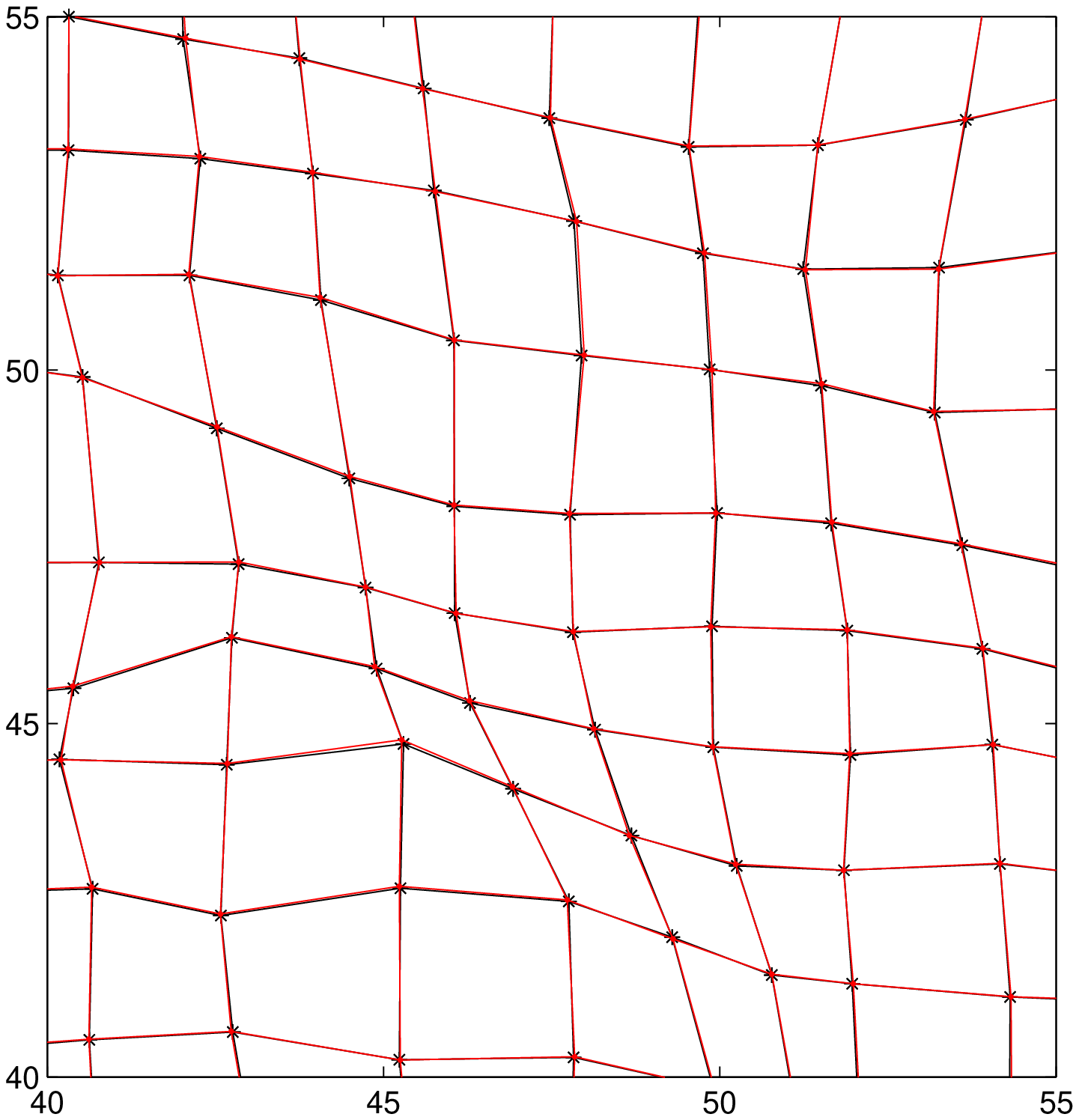}}
\caption{Experiment 1.2, $\alpha=0.1$, $65 \times 65$ grid nodes, 2000 iteration steps, \footnotesize{The black star dots $*$ represent $\mathbf{T}_0$, and red dots $\cdot$ represent constructed $\mathbf{T}$}}
\end{center}
\end{figure}

\begin{table}[H]
\begin{center}
\begin{tabular}{|l|c|c|}\hline
    & \emph{Only Jacobian} & \emph{Jacobian and Curl} \\ \hline
    \emph{$ssd_J$} &0.2640 &0.4783 \\ \hline
    \emph{ssd} & 4.8134 & 1.6869 \\ \hline
    \emph{maximal distance} & 0.3444 & 0.0757 \\ \hline
    \emph{average distance} & 0.1109 & 0.0198 \\ \hline
    \emph{maximal angle difference} & 25.8343 & 16.5480 \\ \hline
    \emph{average angle difference} & 2.6132 & 1.8057 \\ \hline
\end{tabular}
\caption{Comparison of Experiment 1.2}
\end{center}
\end{table}

\begin{figure}[H]
\begin{center}
\subfigure[][Only Jacobian]{\includegraphics[width=0.32\textwidth]{recoverJplusnoisesPhi6Fbrain_1_1.eps}}
\subfigure[][Only Jacobian, enlarged view of left rectangle]{\includegraphics[width=0.32\textwidth]{recoverJplusnoisesPhi6Fbrain_1_2.eps}}
\subfigure[][Only Jacobian, enlarged view of right rectangle]{\includegraphics[width=0.32\textwidth]{recoverJplusnoisesPhi6Fbrain_1_3.eps}}
\subfigure[][Jacobian and Curl]{\includegraphics[width=0.32\textwidth]{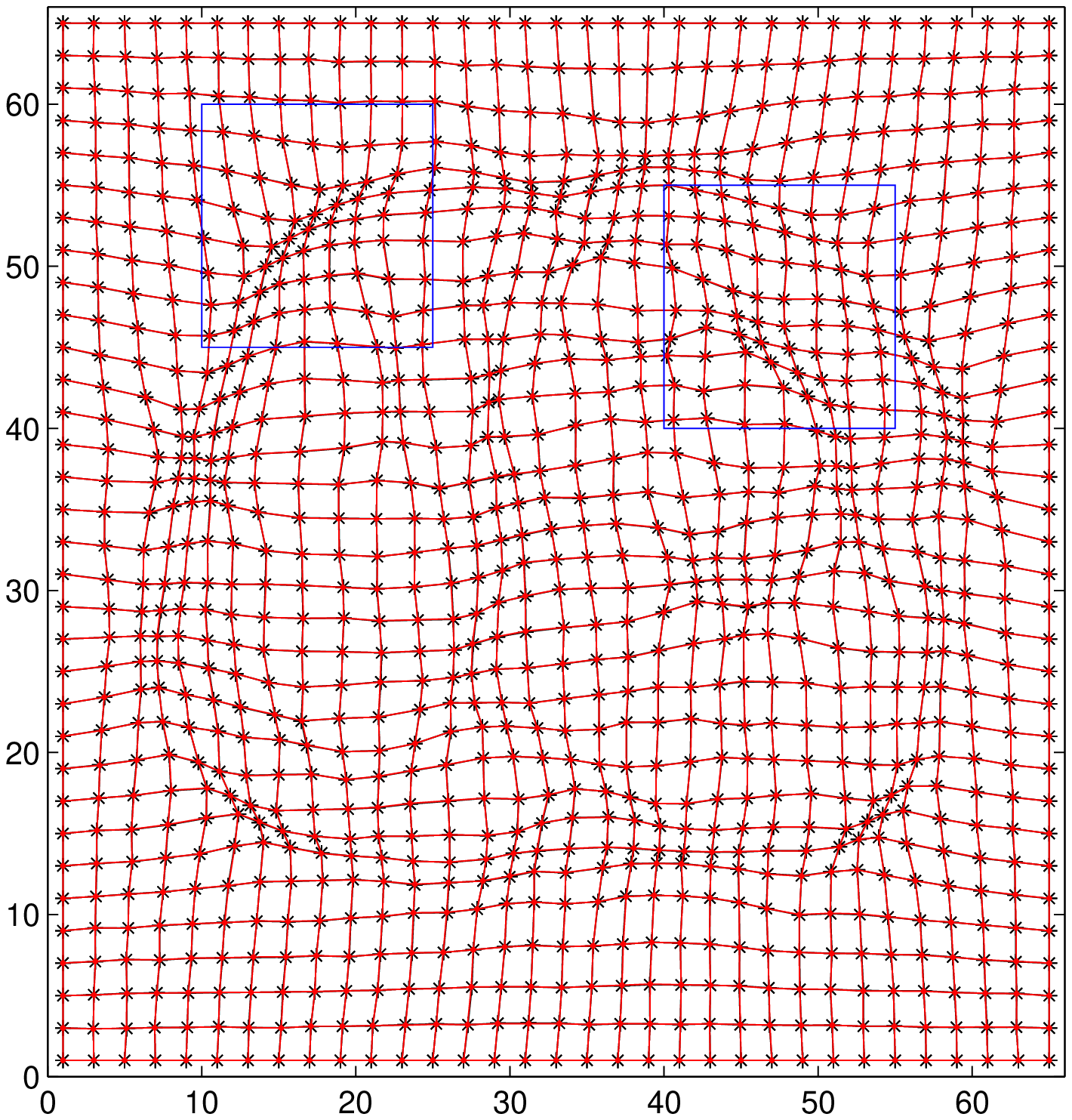}}
\subfigure[][Jacobian and Curl, enlarged view of left rectangle]{\includegraphics[width=0.32\textwidth]{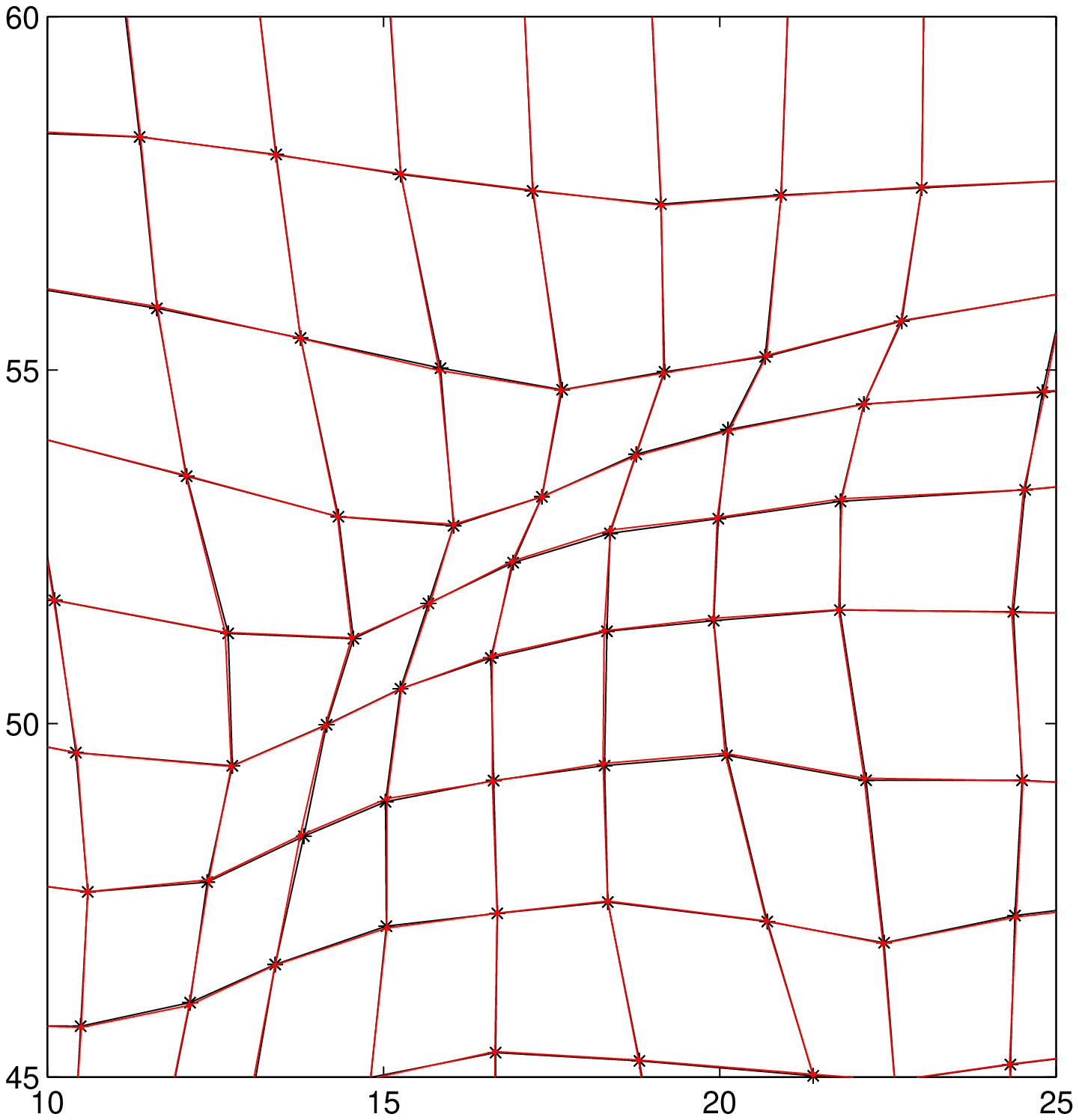}}
\subfigure[][Jacobian and Curl, enlarged view of right rectangle]{\includegraphics[width=0.32\textwidth]{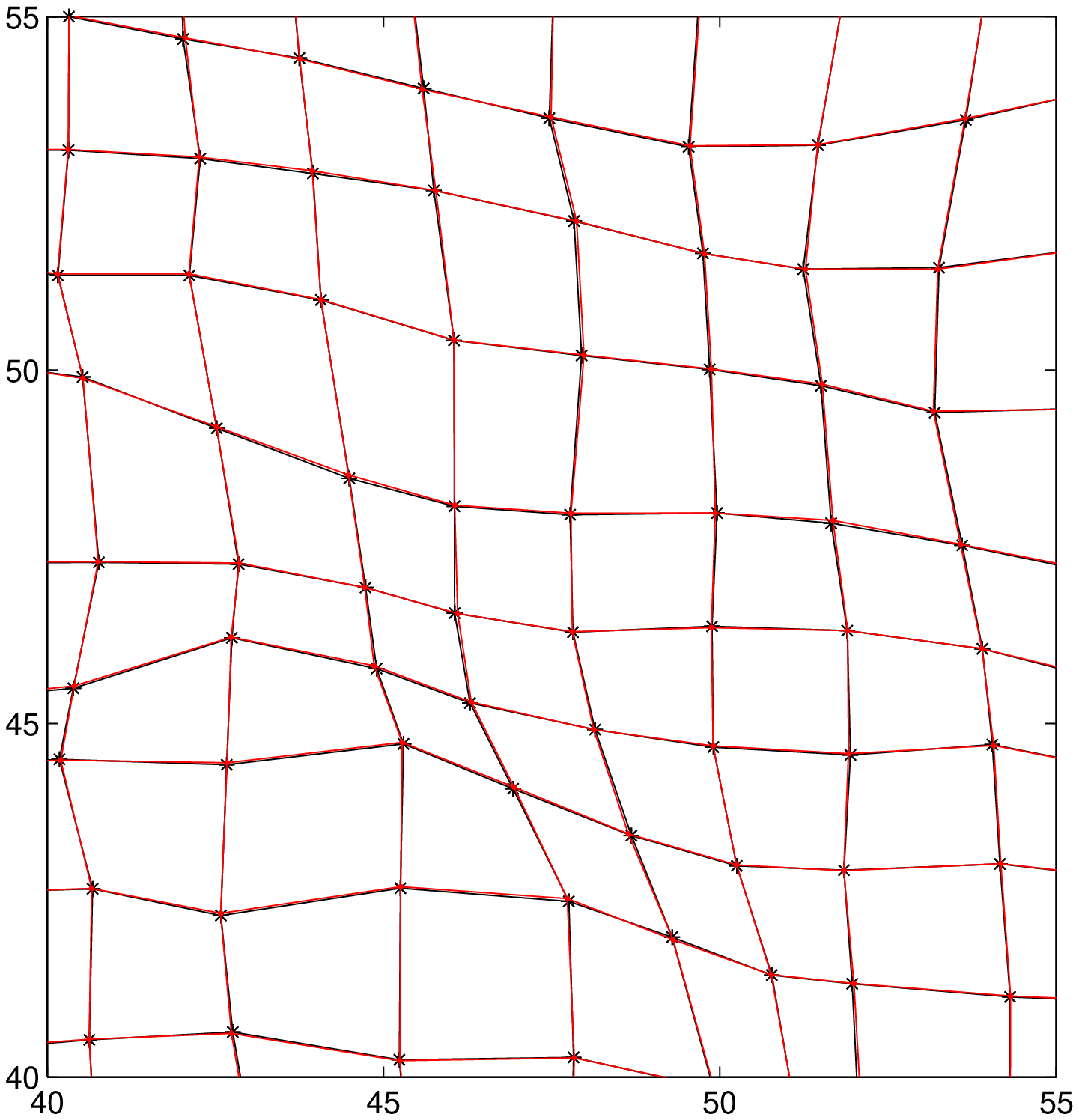}}
\caption{Experiment 1.3, $\alpha=10$, $65 \times 65$ grid nodes, 2000 iteration steps, \footnotesize{The black star dots $*$ represent $\mathbf{T}_0$, and red dots $\cdot$ represent constructed $\mathbf{T}$}}
\end{center}
\end{figure}

\begin{table}[H]
\begin{center}
\begin{tabular}{|l|c|c|}\hline
    & \emph{Only Jacobian} & \emph{Jacobian and Curl} \\ \hline
    \emph{$ssd_J$} &0.2640 &1.4442 \\ \hline
    \emph{ssd} & 4.8134 & 1.8762 \\ \hline
    \emph{maximal distance} & 0.3444 & 0.0797 \\ \hline
    \emph{average distance} & 0.1109 & 0.0203 \\ \hline
    \emph{maximal angle difference} & 25.8343 & 16.6378 \\ \hline
    \emph{average angle difference} & 2.6132 & 1.9603 \\ \hline
\end{tabular}
\caption{Comparison of Experiment 1.3}
\end{center}
\end{table}

\subsubsection{Recover a transformation $\textbf{T}_0$ with moving boundary nodes}
 Now we assume that the transformation $\textbf{T}_0$ has different boundary nodes compared to the background grid. We can use the Remark 1 above to find a boundary match transformation $\textbf{T}^*$ first, then turn back to the fixed boundary nodes case. The following figure 4 shows a experiment for moving boundary nodes case.
\begin{figure}[H]
\begin{center}
\subfigure[][Only Jacobian]{\includegraphics[width=0.32\textwidth]{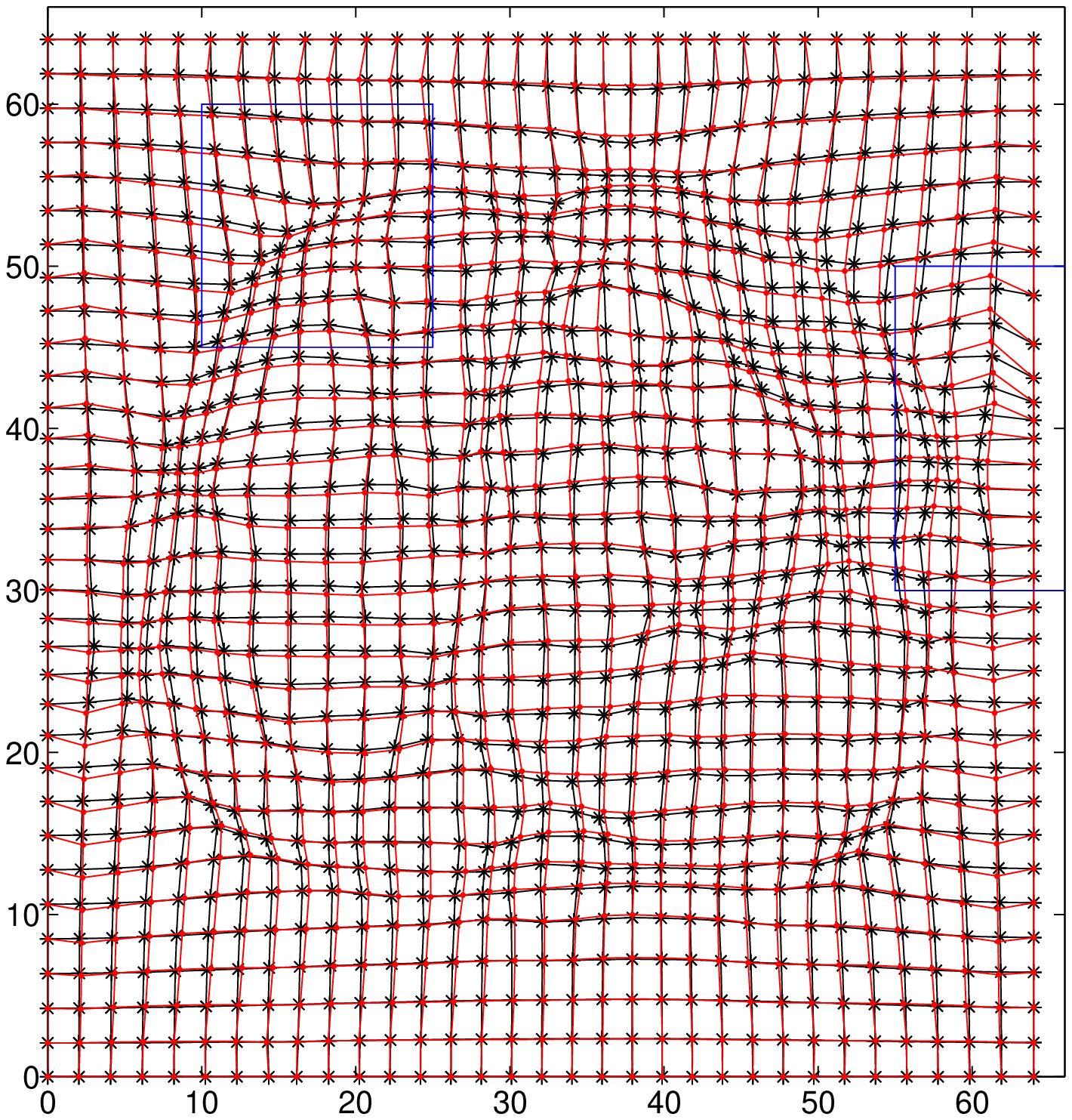}}
\subfigure[][Only Jacobian, enlarged view of left rectangle]{\includegraphics[width=0.32\textwidth]{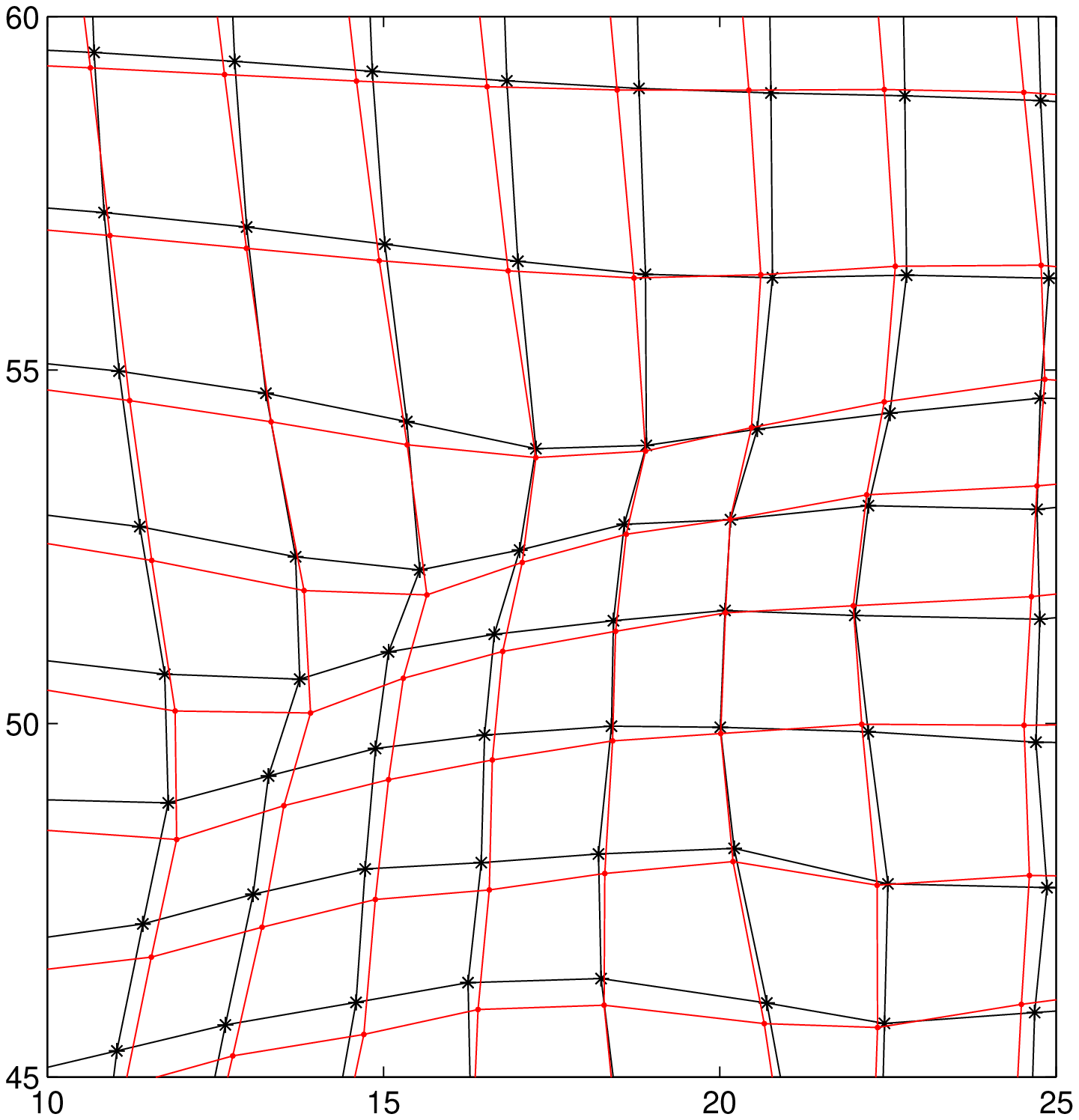}}
\subfigure[][Only Jacobian, enlarged view of right rectangle]{\includegraphics[width=0.32\textwidth]{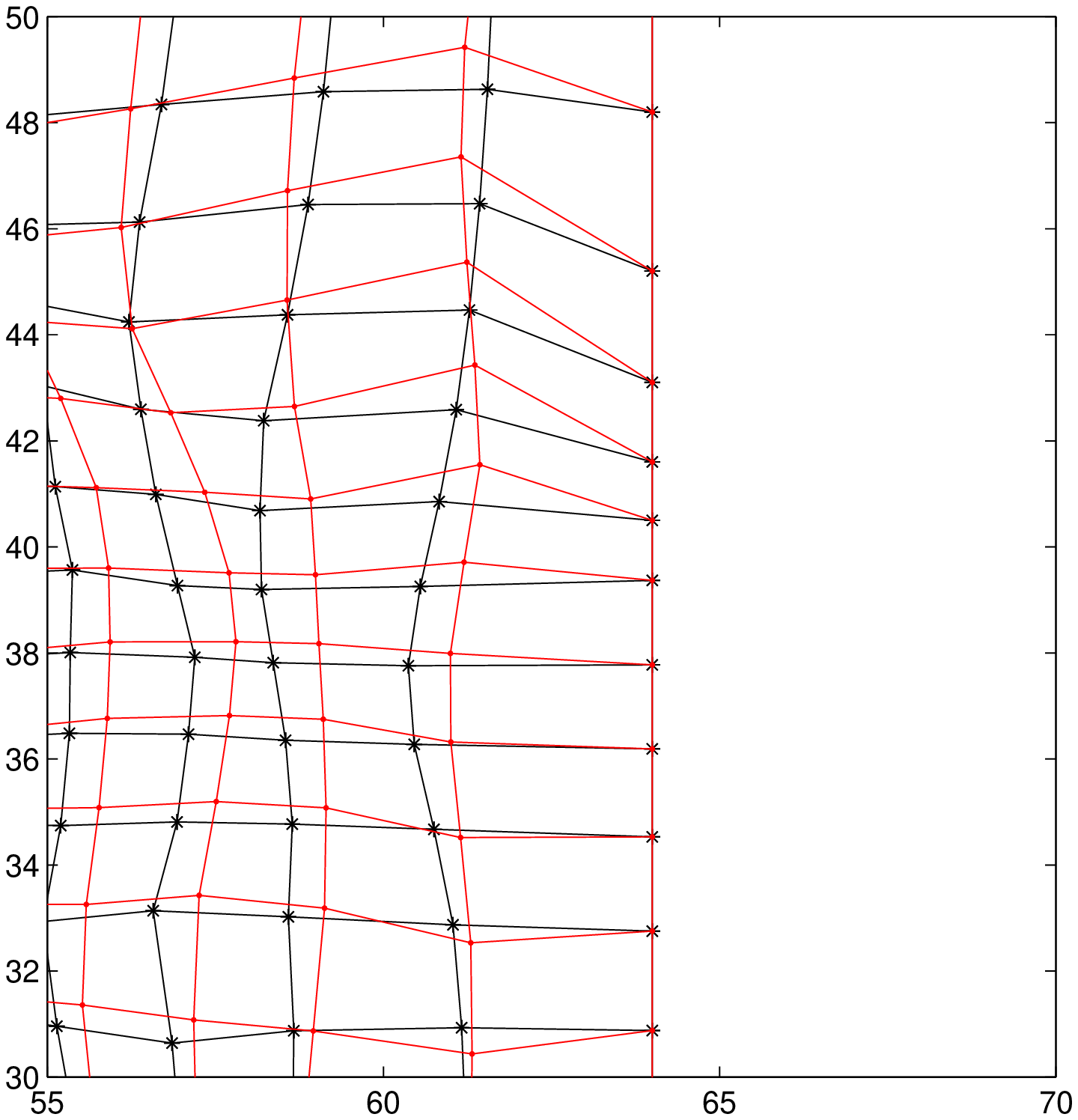}}
\subfigure[][Jacobian and Curl]{\includegraphics[width=0.32\textwidth]{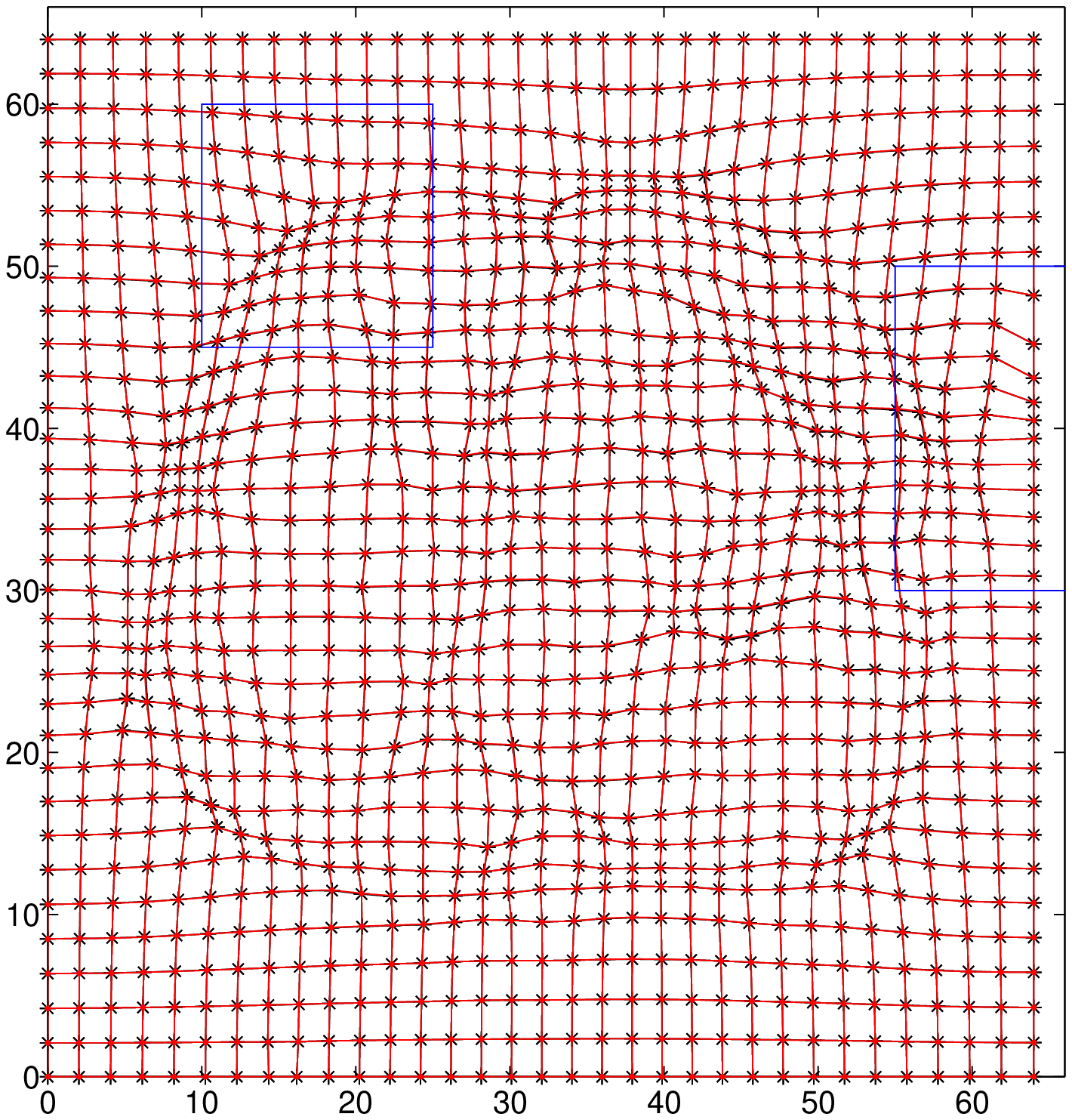}}
\subfigure[][Jacobian and Curl, enlarged view of left rectangle]{\includegraphics[width=0.32\textwidth]{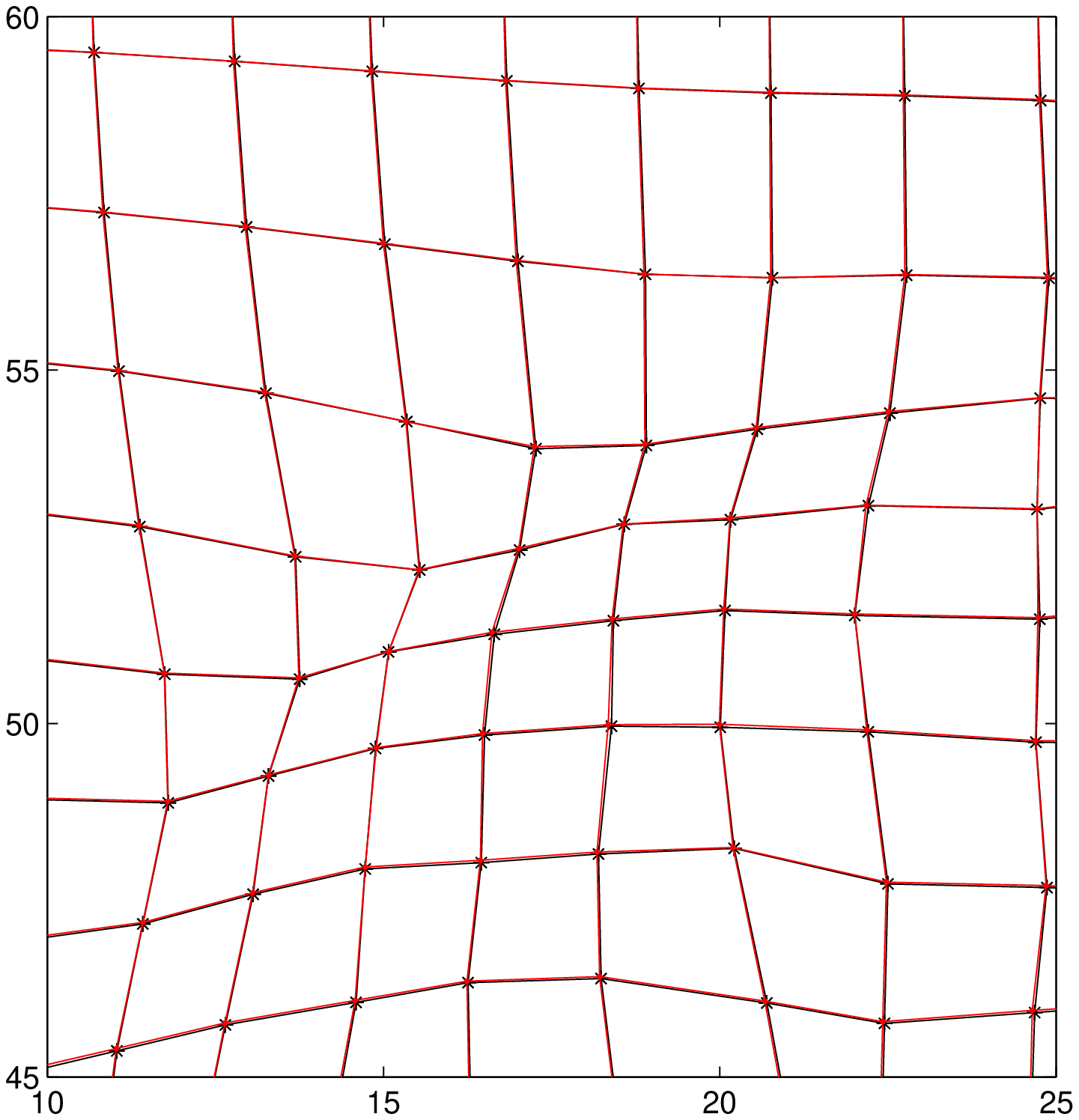}}
\subfigure[][Jacobian and Curl, enlarged view of right rectangle]{\includegraphics[width=0.32\textwidth]{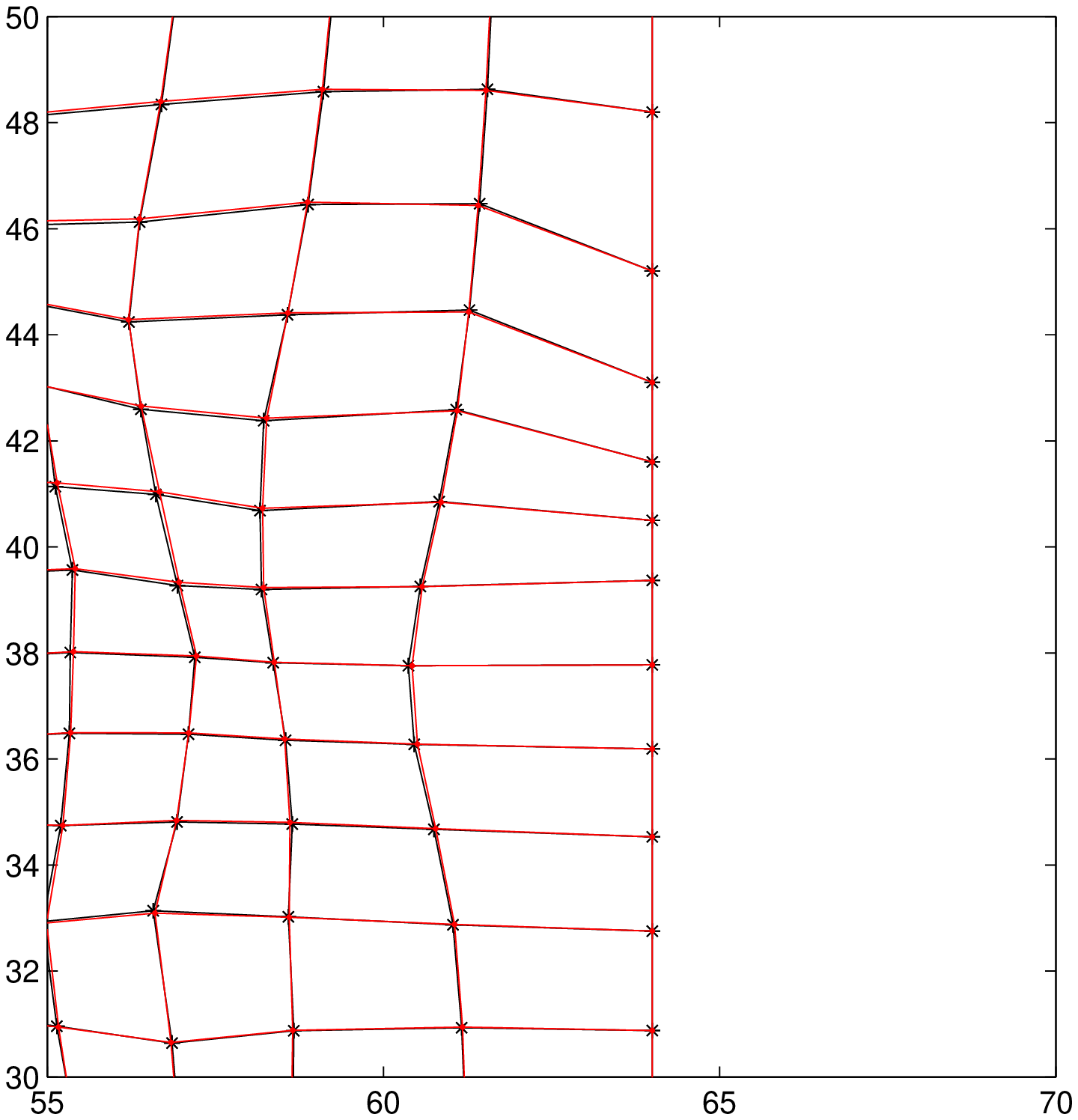}}
\caption{Experiment 2, $\alpha=1$, $65 \times 65$ grid nodes, 2000 iteration steps, \footnotesize{The black star dots $*$ represent $\mathbf{T}_0$, and red dots $\cdot$ represent constructed $\mathbf{T}$}}
\end{center}
\end{figure}

\begin{table}[H]
\begin{center}
\begin{tabular}{|l|c|c|}\hline
    & \emph{Only Jacobian} & \emph{Jacobian and Curl} \\ \hline
    \emph{$ssd_J$} &1.7588 &0.4327 \\ \hline
    \emph{ssd} & 107.8698 & 1.5451 \\ \hline
    \emph{maximal distance} & 1.0158 & 0.1477 \\ \hline
    \emph{average distance} & 0.3316 & 0.0204 \\ \hline
    \emph{maximal angle difference} & 43.4929 & 16.9158 \\ \hline
    \emph{average angle difference} & 4.7968 & 1.2934 \\ \hline
\end{tabular}
\caption{Comparison of Experiment 2}
\end{center}
\end{table}
\section{Conclusion}
We described a new variational method of grid generation with prescribed jacobian determinant and curl, the method has very solid and concise mathematical foundations, related numerical experiment and results also show the method is reliable and accurate. The boundary cases we discussed here almost cover every possible applied case. Also the method can be extended to 3D domain, which will be discussed in further papers.

\end{document}